\documentstyle[prl,aps]{revtex}

\begin{document}
\author{Jian-Qi Shen\footnote{E-mail address: jqshen@coer.zju.edu.cn}}
\address{Zhejiang Institute of Modern Physics and Department of
Physics, Zhejiang University, Hangzhou 310027, P.R. China}
\date{\today}
\title{The su($n$) Lie algebraic structures in the Pegg-Barnett quantization formulation}
\maketitle

\begin{abstract}
 We investigate the oscillator algebra of the
Pegg-Barnett oscillator with a finite-dimensional number-state
space and show that it possesses the su($n$) Lie algebraic
structure. In addition, a so-called supersymmetric Pegg-Barnett
oscillator is suggested, and the related topics such as the
algebraic structure and particle occupation number of the
Pegg-Barnett oscillator are briefly discussed.
\\ \\
{\it PACS:}
\end{abstract}

It is well known that the usual mathematical model of the monomode
quantized electromagnetic field is the harmonic oscillator with an
infinite-dimensional number-state space, the commuting relation of
which is $[a, a^{\dagger}]={\mathcal I}$ with $a$ and
$a^{\dagger}$ being respectively the annihilation and creation
operators of single-mode photon fields. Due to the cyclic
invariance in the trace of the product of two matrices
(operators), {\it i.e.}, ${\rm tr} (aa^{\dagger})={\rm tr}
(a^{\dagger}a)$, it follows directly that the trace of commutator
is vanishing, {\it i.e.}, ${\rm tr}[a, a^{\dagger}]=0$, which,
however, contradicts the fact that the unit matrix ${\mathcal I}$
possesses a nonvanishing trace, namely, ${\rm tr}{\mathcal I}\neq
0$. This, therefore, leads us to consider the oscillator algebra
with finite-dimensional state spaces. On the other hand, in an
attempt to investigate the number-phase uncertainty relations of
the maser and squeezed state in quantum optics, physicists meet,
however, with difficulties arising from a fact that the classical
observable phase of light {\it unexpectedly} has no corresponding
Hermitian operator counterpart (quantum optical
phase)\cite{Louisell,Susskind,Carruthers}. So, several problems we
encountered are as follows: (i) the exponential-form operator
$\exp [i\hat{\phi}]$ (with $\hat{\phi}$ being the phase operator)
is not unitary; (ii) the number-state expectation value of Dirac's
quantum relation $[\hat{\phi}, \hat{N}]=-i$ (with $\hat{N}$ being
the occupation-number operator of photon fields) is even zero,
{\it i.e.}, $\langle n |[\hat{\phi}, \hat{N}]| n\rangle=0$; (iii)
the number-phase uncertainty relation $\Delta N\Delta \phi\geq
\frac{1}{2}$ would imply that a well-defined number state would
actually have a phase uncertainty of greater than $2
\pi$\cite{Pegg}. In order to overcome these difficulties, Pegg and
Barnett suggested an alternative, and physically
indistinguishable, mathematical model of the single-mode field
involving a finite but arbitrarily large state space\cite{Pegg},
in which they defined a phase state as follows
 \begin{equation}
|\theta\rangle=\lim_{s\rightarrow\infty}(s+1)^{-\frac{1}{2}}\sum^{s}_{n=0}\exp
(in\theta)|n\rangle,
\end{equation}
where $|n\rangle$ are the $s+1$ number states, which span an
$(s+1)$ -dimensional state space. This, therefore, means that the
state space $\{|n\rangle\}$ with $0 \leq n\leq s$ has a finitely
upper level ($|s\rangle$) and the maximum occupation number of
particles is $s$ rather than infinity. In their new quantization
formulation, the dimension of number state space is allowed to
tend to infinity after physically measurable results are
calculated\cite{Pegg}. Pegg and Barnett showed that this approach
and the conventional infinite state space are physically
indistinguishable. However, this method has the additional
advantage of being able to incorporate a well-behaved Hermitian
phase operator within the formalism. The resulting number-phase
commutator in Pegg-Barnett approach does not lead to any
inconsistencies yet satisfies the condition for
Poisson-bracket-commutator correspondence. It was shown that
Pegg-Barnett approach has several advantages over the conventional
Susskind-Glogower formulation\cite{Susskind}. For example, the
Pegg-Barnett phase operator is consistent with the vacuum being a
state of random phase, while the Susskind-Glogower phase operator
does not demonstrate such property of the vacuum\cite{Pegg}.
Pegg-Barnett formulation is useful for treating the problems of
atomic coherent population trapping (CPT) and electromagnetically
induced transparency (EIT)\cite{Purdy}.

In this Letter we will further consider the Pegg-Barnett harmonic
oscillator that involves a finitely large state space, and show
that it possesses the su(n) Lie algebraic structures. Based on
this consideration, we will generalize the Pegg-Barnett oscillator
to a supersymmetric one.

The quantum harmonic oscillator possessing an infinite-dimensional
number-state space ({\it i.e.}, the maximum occupation number $s$
tends to infinity) can well model the Bosonic fields. Taking
account of the Pegg-Barnett (P-B) harmonic oscillator means that
the non-semisimple Lie algebra should be generalized to the
semisimple one, namely, we will replace the familiar quantum
commutator $[a, a^{\dagger}]={\mathcal I}$ with the new $[a,
a^{\dagger}]={\mathcal A}$ (${\mathcal A}$ will be defined in what
follows). For a preliminary consideration, we take into account
the case $s=1$, where the matrix representation of the
annihilation (creation) operators and ${\mathcal A}$ of the fields
are of the form (in the number-state basis set)
\begin{equation}
a=\left(\begin{array}{cccc}
0  & 1  \\
0 &   0  \\
 \end{array}
 \right),                 \quad         a^{\dagger}=\left(\begin{array}{cccc}
0 & 0 \\
1 & 0  \\
 \end{array}
 \right),                     \quad         {\mathcal A}=\left(\begin{array}{cccc}
1 & 0 \\
0 & -1  \\
 \end{array}
 \right).
 \label{eq1}
\end{equation}
It is apparently seen that the operators $a$, $a^{\dagger}$ and
${\mathcal A}$ satisfy an su($2$) algebraic commuting relations.
Here one can readily verify that
$a=\frac{\sigma_{1}+i\sigma_{2}}{2}$,
$a^{\dagger}=\frac{\sigma_{1}-i\sigma_{2}}{2}$ and ${\mathcal
A}=\sigma_{3}$, where $\sigma_{i}$'s ($i=1,2,3$) are Pauli's
matrices. It follows from (\ref{eq1}) that
$aa^{\dagger}+a^{\dagger}a={\mathcal I}$. This, therefore, implies
that the P-B harmonic oscillator with $s=1$ corresponds to the
Fermionic fields and possesses the su($2$) Lie algebraic
structure.

In what follows we will study the algebraic structures of P-B
harmonic oscillators with arbitrary occupation number $s$.  As
another illustrative example, here we will take into consideration
the case of $s=2$, the matrix representation of $a$, $a^{\dagger}$
and ${\mathcal A}$ of which are written (in the number-state basis
set)
\begin{equation}
a=\left(\begin{array}{cccc}
0 & 1 & 0 \\
0 &   0 & \sqrt{2}  \\
 0 &  0 &  0
 \end{array}
 \right),                 \quad          a^{\dagger}=\left(\begin{array}{cccc}
0  & 0 & 0 \\
1 &   0 & 0  \\
 0 &  \sqrt{2} &  0
 \end{array}
 \right),                 \quad          {\mathcal A}=\left(\begin{array}{cccc}
1  & 0 & 0 \\
0 &   1 & 0  \\
 0 &  0 &  -2
 \end{array}
 \right).
 \label{eq2}
\end{equation}
Calculation of the commutators among the Lie algebraic generators
of the P-B harmonic oscillator with $s=2$ yields\footnote{This
work was finished in 2000.}
\begin{eqnarray}
\left[a,  {\mathcal A}\right]&=&3\sqrt{2}\left(\begin{array}{cccc}
0  & 0 & 0 \\
0&  0 &-1  \\
 0 &  0 &  0
 \end{array}
 \right)=3\sqrt{2}{\mathcal M},      \quad          \quad     [a^{\dagger},  {\mathcal A}]
 =-3\sqrt{2}\left(\begin{array}{cccc}
0  & 0 & 0 \\
0&  0 &0  \\
 0 &  -1 &  0
 \end{array}
 \right)=-3\sqrt{2}{\mathcal M}^{\dagger},             \nonumber \\
\left[{\mathcal M}, {\mathcal M}^{\dagger}\right]
&=&-\left(\begin{array}{cccc}
0  & 0 & 0 \\
0&  -1 &0  \\
0 &  0&  1
\end{array}
\right)=-{\mathcal K},                    \quad          \quad
                         \left[a,  {\mathcal M}\right]=-\left(\begin{array}{cccc}
0  & 0 & 1 \\
0&  0 &0  \\
 0 &  0 &  0
 \end{array}
 \right)=-{\mathcal F},
                \nonumber \\
 \left[a^{\dagger}, {\mathcal M}\right]&=&-\sqrt{2}{\mathcal K},
        \quad      \left[a, {\mathcal M}^{\dagger}\right]=\sqrt{2}{\mathcal K},
           \quad       \left[a^{\dagger}, {\mathcal M}^{\dagger}\right]={\mathcal F}^{\dagger},
                 \quad     [{\mathcal K}, {\mathcal F}]=-{\mathcal
                 F},  \quad      [{\mathcal K}, {\mathcal F}^{\dagger}]={\mathcal
                 F}^{\dagger}, \quad  ...    \label{eq3}
\end{eqnarray}
Further calculation shows that the algebraic generators $a,
a^{\dagger}, {\mathcal A}, {\mathcal M}, {\mathcal M}^{\dagger},
{\mathcal K}, {\mathcal F}, {\mathcal F}^{\dagger}$ form the
su($3$) algebra, since the eight Gell-Mann matrices can be
constructed in terms of them, {\it i.e.},
\begin{eqnarray}
\lambda_{1}&=&a+a^{\dagger}+\sqrt{2}({\mathcal M}+{\mathcal
M}^{\dagger}),   \quad
\lambda_{2}=i[a^{\dagger}-a+\sqrt{2}({\mathcal
M}^{\dagger}-{\mathcal M})], \quad     \lambda_{3}={\mathcal
A}+2{\mathcal K},  \quad    \lambda_{4}={\mathcal F}+{\mathcal
F}^{\dagger},
                 \nonumber \\
    \lambda_{5}&=&i({\mathcal
F}^{\dagger}-{\mathcal F}),   \quad \lambda_{6}=-({\mathcal
M}+{\mathcal M}^{\dagger}), \quad \lambda_{7}=-i({\mathcal
M}^{\dagger}-{\mathcal M}),     \quad
\lambda_{8}=\frac{1}{\sqrt{3}}\lambda_{8}.
\end{eqnarray}
Thus we show that the P-B harmonic oscillator with $s=2$ possesses
the su($3$) Lie algebraic structure.

For the P-B harmonic oscillator with a finite but arbitrarily
large state space of $s+1$ dimensions, the matrix representation
(in the number-state basis set) of the operators $a$,
$a^{\dagger}$ and ${\mathcal A}$ takes the following form
\begin{equation}
a_{mn}=\sqrt{n}\delta_{m, n-1},  \quad
a^{\dagger}_{mn}=\sqrt{n+1}\delta_{m, n+1},   \quad    {\mathcal
A}_{mn}=\delta_{mn}-(s+1)\delta_{ms}\delta_{ns},
\end{equation}
where the subscript $m, n$ denote the matrix indices. The
remaining generators ${\mathcal M}, {\mathcal M}^{\dagger},
{\mathcal K}, {\mathcal F}, {\mathcal F}^{\dagger}, ...$ can be
obtained as follows ($0\leq m, n\leq s$):
\begin{eqnarray}
\left[a, {\mathcal A}\right]_{mn}&=&(s+1)\sqrt{s}(-\delta_{m+1,
s}\delta_{n
s})=(s+1)\sqrt{s}{\mathcal M}_{mn},   \nonumber \\
\left[a^{\dagger}, {\mathcal
A}\right]_{mn}&=&-(s+1)\sqrt{s}(-\delta_{m s}\delta_{n+1,
s})=-(s+1)\sqrt{s}{\mathcal M}^{\dagger}_{mn},
              \nonumber \\
\left[{\mathcal M}, {\mathcal
M}^{\dagger}\right]_{mn}&=&-(\delta_{ms}\delta_{ns}-\delta_{m+1,s}\delta_{n+1,s})=-{\mathcal
K}_{mn},
\nonumber \\
\left[ {\mathcal A}, {\mathcal M}\right]&=&(1+s){\mathcal M},
\quad           \left[ {\mathcal A}, {\mathcal
M}^{\dagger}\right]=-(1+s){\mathcal M}^{\dagger},
\nonumber \\
\left[a, {\mathcal M}\right]_{mn}&=&-\sqrt{s-1}\delta_{m+1,
s-1}\delta_{ns}=-\sqrt{s-1}{\mathcal F}_{mn},
\nonumber \\
\left[a^{\dagger}, {\mathcal
M}^{\dagger}\right]_{mn}&=&\sqrt{s-1}\delta_{ms}\delta_{n+1,
s-1}=\sqrt{s-1}{\mathcal F}^{\dagger}_{mn}, \nonumber  \\
\left[{\mathcal K}, {\mathcal F}\right]&=&-{\mathcal F}, \quad
\left[{\mathcal K}, {\mathcal F}^{\dagger}\right]={\mathcal
F}^{\dagger},   \quad    \left[{\mathcal M}, {\mathcal
K}\right]=2{\mathcal M},  \quad             \left[{\mathcal
M}^{\dagger}, {\mathcal K}\right]=-2{\mathcal M}^{\dagger}, \quad
...                      \label{eq4}
\end{eqnarray}
For the case of $s=2$, it has been shown above that Hermitian
operators (such as the eight Gell-Mann matrices) can be
constructed in terms of $a, a^{\dagger}, {\mathcal A}, {\mathcal
M}, {\mathcal M}^{\dagger}, {\mathcal K}, {\mathcal F}, {\mathcal
F}^{\dagger}$. Likewise, here the Hermitian operators (generators)
of Lie algebra can also be obtained via the linear combination of
the above generators (\ref{eq4}). If ${\mathcal G}$ represents the
linear combination of the Hermitian operators, and consequently
${\mathcal G}={\mathcal G}^{\dagger}$, then the exponential-form
group element operator $U=\exp (i{\mathcal G})$ is unitary.
Besides, since $a$, $a^{\dagger}$ and ${\mathcal A}$ are
traceless, all the generators derived by the commutators in
(\ref{eq4}) (and hence ${\mathcal G}$) are also traceless due to
the cyclic invariance in the trace of matrices product. Thus the
determinant of the group element $U$ is unit, {\it i.e.}, ${\rm
det}U=1$, because of ${\rm det}U=\exp [{\rm tr}(i{\mathcal G})]$.
Since it is known that such $U$ that satisfies simultaneously the
above two conditions is the group element of the su($n$) Lie
group, the above-presented generators $a, a^{\dagger}, {\mathcal
A}, {\mathcal M}, {\mathcal M}^{\dagger}, {\mathcal K}, {\mathcal
F}, {\mathcal F}^{\dagger}, ...$ will close corresponding su($n$)
Lie algebraic commutation relations among themselves. It is thus
concluded that the P-B harmonic oscillator with the maximum
occupation number being $s$ has an $(s+1)$- dimensional
number-state space and possesses the su($s+1$) Lie algebraic
structure.

Considering the case of $s\rightarrow \infty$ is of physically
typical interest. Apparently, it is seen that ${\mathcal A}$ tends
to a unit matrix ${\mathcal I}$, and the off-diagonal matrix
elements of all other generators except $a, a^{\dagger}$ approach
the zero matrices ${\mathcal O}$. This, therefore, means that the
P-B harmonic oscillator with infinite-dimensional state space just
corresponds to the Bosonic fields.
\\ \\

In conclusion, in the above we extend the non-semisimple algebra
of harmonic oscillator with infinite-dimensional state space to a
semisimple algebraic case, which can characterize the algebraic
structures of the P-B oscillator. In what follows we will consider
a generalization of P-B oscillator, {\it i.e.}, the so-called
supersymmetric  Pegg-Barnett oscillator, which may possess some
physically interesting significance.

 For this aim, we will take into account a set of algebraic generators
$(N,N^{^{\prime }},Q,Q^{\dagger})$ which possesses a
supersymmetric Lie algebraic properties, {\it i.e.},

\begin{eqnarray}
Q^{2} &=&(Q^{\dagger })^{2}=0,\quad \left[ Q,Q^{\dagger}\right]
=N^{^{\prime }}\sigma _{z},\quad \left[ N,N^{^{\prime }}\right] =0,\quad %
\left[ N,Q\right] =-Q,  \nonumber \\
\left[ N,Q^{\dagger }\right] &=&Q^{\dagger },    \quad \left\{
Q,Q^{\dagger }\right\} =N^{^{\prime }},     \quad \left\{ Q,\sigma
_{z}\right\} =\left\{
Q^{\dagger },\sigma _{z}\right\} =0,  \nonumber \\
\left[ Q,\sigma _{z}\right] &=&-2Q,      \quad
 \left[ Q^{\dagger },\sigma _{z}
\right] =2Q^{\dagger },              \quad                 \left(
Q^{\dagger }-Q\right) ^{2}=-N^{^{\prime }},  \label{eq33}
\end{eqnarray}
where $\left\{ {}\right\} $ denotes the anticommuting bracket.
Such Lie algebra (\ref{eq33}) can be physically realized by the
two-level multiphoton Jaynes-Cummings model, the Hamiltonian
(under the rotating wave approximation) of which is of the
form\cite{Sukumar,Kien,Shen2}

\begin{equation}
H=\omega a^{\dagger }a+\frac{\omega _{0}}{2}\sigma
_{z}+g(a^{\dagger })^{k}\sigma _{-}+g^{\ast }a^{k}\sigma _{+},
\label{eq31}
\end{equation}
where $a^{\dagger }$ and $a$ are the creation and annihilation
operators for the electromagnetic field, and obey the commutation
relation $\left[ a,a^{\dagger }\right] =1$; $\sigma _{\pm }$ and
$\sigma _{z}$ denote the two-level atom operators which satisfy
the commutation relation $\left[ \sigma _{z},\sigma _{\pm }\right]
=\pm 2\sigma _{\pm }$ ; $g$ and $ g^{\ast }$ are the coupling
coefficients and $k$ is the photon number in each atom transition
process; $\omega _{0}$ and $\omega$ represent respectively the
transition frequency and the mode frequency. By the aid of
(\ref{eq33}) and the following expressions
(\ref{eq32})\cite{Lu1,Lu2,Shen1}
\begin{eqnarray}
N &=&a^{\dagger }a+\frac{k-1}{2}\sigma _{z}+\frac{1}{2}=\left(
\begin{array}{cc}
a^{\dagger }a+\frac{k}{2} & 0 \\
0 & aa^{\dagger }-\frac{k}{2}
\end{array}
\right) ,                             \quad N^{^{\prime }}=\left(
\begin{array}{cc}
\frac{a^{k}(a^{\dagger })^{k}}{k!} & 0 \\
0 & \frac{(a^{\dagger })^{k}a^{k}}{k!}
\end{array}
\right) ,  \nonumber \\
Q^{\dagger } &=&\frac{1}{\sqrt{k!}}(a^{\dagger })^{k}\sigma
_{-}=\left(
\begin{array}{cc}
0 & 0 \\
\frac{(a^{\dagger })^{k}}{\sqrt{k!}} & 0
\end{array}
\right) ,              \quad
 Q=\frac{1}{\sqrt{k!}}a^{k}\sigma
_{+}=\left(
\begin{array}{cc}
0 & \frac{a^{k}}{\sqrt{k!}} \\
0 & 0
\end{array}
\right) ,  \label{eq32}
\end{eqnarray}
the Hamiltonian (\ref{eq31}) of the two-level multiphoton
Jaynes-Cummings model can be rewritten as

\begin{equation}
H=\omega N+\frac{\omega -\delta }{2}\sigma _{z}+g Q^{\dagger
}+g^{\ast }Q-\frac{\omega}{2}         \label{eq34}
\end{equation}
with $\delta =k\omega -\omega _{0}$.

The present illustrative example (the supersymmetric multiphoton
Jaynes-Cummings model) is helpful for understanding the physical
meanings of above supersymmetric algebra. But note that the
concept of supersymmetric P-B oscillator that will be put forward
in the following is not related to the above multiphoton
Jaynes-Cummings model (\ref{eq31}) at all. Use is made of
$\frac{1}{k!}a^{k}(a^{\dagger })^{k}\left| m\right\rangle
=\frac{(m+k)!}{m!k!}\left| m\right\rangle $ and
$\frac{1}{k!}(a^{\dagger })^{k}a^{k}\left| m+k\right\rangle
=\frac{(m+k)!}{m!k!}\left| m+k\right\rangle$, and then one can
arrive at

\begin{equation}
N^{^{\prime }}{%
{\left| m\right\rangle  \choose \left| m+k\right\rangle }%
}=C_{m+k}^{m}{
{\left| m\right\rangle  \choose \left| m+k\right\rangle }
}  \label{eq35}
\end{equation}
with $C_{m+k}^{m}=\frac{(m+k)!}{m!k!}.$ Thus we obtain the
supersymmetric quasialgebra $(N,Q,Q^{\dagger },\sigma _{z})$ in
the sub-Hilbert-space corresponding to the particular eigenvalue
$C_{m+k}^{m}$ of the Lewis-Riesenfeld invariant operator
$N^{^{\prime }}$\cite{Shen1} by replacing the generator
$N^{^{\prime }}$ with $C_{m+k}^{m}$ in the commutation relations
in (\ref{eq33}), namely,

\begin{equation}
\left[ Q,Q^{\dagger }\right] =C_{m+k}^{m}\sigma _{z}, \quad
\left\{ Q,Q^{\dagger }\right\} =C_{m+k}^{m},         \quad \left(
Q^{\dagger }-Q\right) ^{2}=-C_{m+k}^{m}.  \label{eq36}
\end{equation}

Based on the discussion of such quasialgebra in the
sub-Hilbert-space corresponding to the particular eigenvalue
$C_{m+k}^{m}$ of the generator $N^{^{\prime }}$, we can propose
the supersymmetric P-B oscillator. The algebraic generators of the
generalized P-B oscillator under consideration agree with the
commutation relation (\ref{eq36}), where $Q^{\dagger}$ and $Q$ can
be regarded as the creation and annihilation operators and the
eigenvalue $C_{m+k}^{m}$ of $N'$ may be considered the particle
occupation number of the P-B oscillator in a certain number state.
Evidently, if $k=0$, then the supersymmetric P-B oscillator is
reduced to the regular Fermionic case.
\\ \\

It is believed that the extension of P-B formulation to the
supersymmetric case may be physically interesting. For example, we
have studied the mass spectrum of the charged leptons and obtained
the following mass formula\footnote{This mass formula was
suggested in November 1999. It has, however, never been published
elsewhere yet up to now.}

\begin{equation}
m_{n}=C_{3}^{n}\left(\frac{1}{2}\right)^{n^{2}}\left(\frac{1}{\alpha}\right)^{n}m_{\rm
e}                  \label{eq37}
\end{equation}
with $m_{\rm e}$ and $\alpha$ being the electron mass and the
electromagnetic fine structure constant, respectively. The integer
$n$ in (\ref{eq37}) denotes the various generations of charged
leptons, {\it i.e.}, the electron, muon ($\mu$) and tau ($\tau$)
particle correspond to $n=0, 1, 2$, respectively. The present mass
formula (\ref{eq37}) agrees to the experimental results about one
part in $10^{3}$. It follows from the charged leptons mass formula
(\ref{eq37}) that the maximum $n$ can take $n=3$ and the total
generation number of charged leptons may therefore be $4$, and
hence there might exist a fourth charged lepton in nature, which
is unknown up to now. According to (\ref{eq37}), the mass of the
potential charged lepton, which we call, for brevity, f lepton, is
about $5022$ times that of electron. The name of this hypothetical
charged lepton is inspired by considering that it may be the {\it
fourth}- (or even {\it final}-) generation charged lepton.

As far as the generalized P-B oscillator is concerned, the
algebraic commutation relation (\ref{eq36}) may clue physicists on
the mathematical mechanism and physical meanings of the above mass
spectrum of charged leptons. Even though at present it is well
known that various experimental evidences show that there are only
three generations of fundamental particles\cite{Dolgov}, the
detection of potential new generation of particles is still of
physical interest. This subject is beyond the scope of the present
Letter and will be published elsewhere.
\\ \\

To summarize, in this Letter we consider the su($n$) Lie algebraic
structure in the P-B quantization formulation and generalize the
P-B oscillator to the supersymmetric case. It is shown that the
Fermionic and Bosonic fields are two special cases of P-B
oscillator, the corresponding dimensions of state spaces of which
are 2 and infinity, respectively. The potential application of the
supersymmetric P-B oscillator algebra to the mass spectrum of
charged leptons is briefly suggested. We hope the consideration of
algebraic structures of P-B oscillator presented here may open up
new opportunities for investigating the generation number of
particles (should such f lepton exist) and other related topics
such as fractional statistics, anyon\cite{Wilczek} and cyclic
representation of quantum algebra/group\cite{Fujikawa}.
\\ \\

\textbf{Acknowledgements}  This project was supported partially by
the National Natural Science Foundation of China under the project
No. $90101024$.


\begin{references}
\bibitem{Louisell} W.H. Louisell, Phys. Lett. 7 (1963) 60.
\bibitem{Susskind} L. Susskind, J. Glogower, Physics 1 (1964) 49.
\bibitem{Carruthers} P. Susskind, M. M. Nieto, Rev. Mod. Phys. 40
(1968) 411.
\bibitem{Pegg} D.T. Pegg, S. M. Barnett, Phys. Rev. A 39 (1989)
1665.
\bibitem{Purdy} T. Purdy, M. Ligare, J. Opt. B 5 (2003) 289.

\bibitem{Sukumar}  C.V. Sukumar, B. Buck, Phys. Lett. A  83 (1981) 211.

\bibitem{Kien}  F.L. Kien, M. Kozierowki, T. Quany, Phys. Rev. A 38
(1988) 263.

\bibitem{Shen2}   J.Q. Shen, H.Y. Zhu, H. Mao, J. Phys. Soc.
Jpn. 71 (2002) 1440.

\bibitem{Lu1}  H.X. Lu, X.Q. Wang, Y.D. Zhang, Chin. Phys. 9 (2000)
325

\bibitem{Lu2}  H.X. Lu, X.Q. Wang, Chin. Phys. 9 (2000) 568

\bibitem{Shen1} J.Q. Shen, H.Y. Zhu, P. Chen, Euro. Phys. J. D 23 (2003) 305.

\bibitem{Dolgov} A.D. Dolgov, Y.B. Zeldovich, Rev. Mod. Phys. 53
(1981) 1.

\bibitem{Wilczek} F. Wilczek, Phys. Rev. Lett. 48 (1982) 114; 49 (1982) 957.

\bibitem{Fujikawa} K. Fujikawa, Phys. Rev. A 52 (1995) 3299.

\end{references}
\end{document}